\input harvmac
\input amssym.def
\input amssym.tex

\def\tablerule{\noalign{\hrule}}
\parskip=4pt \baselineskip=12pt
\hfuzz=20pt
\parindent 10pt

\def\ta{{\tilde\alpha}}

\def\nt{\noindent}
\def\nl{\hfill\break}

\def\np{\vfill\eject}

\global\newcount\subsubsecno \global\subsubsecno=0
\def\subsubsec#1{\global\advance\subsubsecno
by1\message{(\secsym\the\subsecno.\the\subsubsecno. #1)}
\ifnum\lastpenalty>9000\else\bigbreak\fi
\noindent{\bf\secsym\the\subsecno.\the\subsubsecno.
#1}\writetoca{\string\quad {\secsym\the\subsecno.\the\subsubsecno.}
{#1}}\par\nobreak\medskip\nobreak}

\def\subsubsec#1{\global\advance\subsubsecno
by1\message{(\secsym\the\subsecno.\the\subsubsecno. #1)}
\ifnum\lastpenalty>9000\else\bigbreak\fi
\noindent{\bf\secsym\the\subsecno.\the\subsubsecno.
#1}\writetoca{\string\quad
{\secsym\the\subsecno.\the\subsubsecno.} {#1}}}%

\input epsf.tex
\newcount\figno
\figno=0
\def\fig#1#2#3{
\par\begingroup\parindent=0pt\leftskip=1cm\rightskip=1cm\parindent=0pt
\baselineskip=11pt \global\advance\figno by 1 \midinsert
\epsfxsize=#3 \centerline{\epsfbox{#2}} \vskip 12pt
#1\par
\endinsert\endgroup\par}
\def\figlabel#1{\xdef#1{\the\figno}}
\def\encadremath#1{\vbox{\hrule\hbox{\vrule\kern8pt\vbox{\kern8pt
\hbox{$\displaystyle #1$}\kern8pt} \kern8pt\vrule}\hrule}}

\def\tN{{\tilde N}}  \def\tV{{\tilde V}}
\def\tcn{{\tilde{\cal N}}}

\def\rank{{\rm rank}}
\def\downcirc#1{\mathop{\circ}\limits_{#1}}
\def\riga{-\kern-4pt - \kern-4pt -}
\font\fat=cmsy10 scaled\magstep5

\def\Bbullet{\raise-3pt\hbox{\fat\char"0F}}

\def\black#1{\mathop{\bullet}\limits_{#1}}

\font\sfont=cmbx12 scaled\magstep1 
\font\tfont=cmbx12 scaled\magstep2 

\font\male=cmr9

\def\Box{
\vbox{ \halign to5pt{\strut##& \hfil ## \hfil \cr &$\kern -0.5pt
\sqcap$ \cr \noalign{\kern -5pt \hrule} }}~}

\def\down{\raise1.5pt\hbox{$\phantom{a}_2$}\downarrow}

\def\downa{\raise1.5pt\hbox{$\phantom{a}_{2\atop m_2}$}\downarrow}

\def\llr{\longrightarrow}

\def\({\left(}
\def\){\right)}

\def\lra{\longrightarrow}

\def\dia{{$\diamondsuit$}}

\def\ha{{\textstyle{1\over2}}}

\def\pd{\partial}

\def\bbc{{C\kern-6.5pt I}}
\def\bac{{C\kern-5.5pt I}}
\def\bab{{C\kern-4.5pt I}}

\def\bbr{{I\!\!R}}
\def\bbn{I\!\!N}
\def\a{\alpha}
\def\b{\beta}

\def\vr{\vert}

\def\l{\lambda}

\def\ca{{\cal A}}  \def\cc{{\cal C}}
\def\cd{{\cal D}} \def\ce{{\cal E}} \def\cf{{\cal F}}
\def\cg{{\cal G}} \def\ch{{\cal H}} 
 \def\ck{{\cal K}} 
\def\cm{{\cal M}} \def\cn{{\cal N}} 
\def\cp{{\cal P}}  
 \def\ct{{\cal T}}

\def\tn{{\tilde n}}

\def\ido{intertwining differential operator}
\def\idos{intertwining differential operators}

\def\L{\Lambda}
\def\r{\rho}

\def\tZ{{\tilde Z}}


\nref\KanLan{J. Kang and P. Langacker, Phys. Rev. {\bf D71} (2005)
035014, hep-ph/0412190.}

\nref\GanVas{T. Gannon and M. Vasudevan, JHEP 0507:035 (2005),
hep-th/0504006.}

\nref\Ohwa{Y. Ohwashi, Prog. Theor. Phys. {\bf 115}, 625-659 (2006),
 hep-th/0510252.}

\nref\Tuite{M.P. Tuite,
 math/0610322, arXiv:0811.4523 [math.QA].}

\nref\KelRos{C.A. Keller and S. Rossi,
  JHEP 0703:038 (2007),  hep-th/0610175.}

\nref\HunMos{P.Q. Hung and Paola Mosconi,
$E_6$ unification in a model of dark energy and dark matter,
hep-ph/0611001.}

\nref\DufFer{M.J. Duff and S. Ferrara, Phys. Rev. {\bf D76}, 124023
(2007), arXiv:0704.0507v2 [hep-th].}

\nref\FaKoRi{A.E. Faraggi, C. Kounnas and J. Rizos, Nucl. Phys. {\bf
B774}, 208-231 (2007), hep-th/0611251.}

\nref\KiMoNe{S.F. King, S. Moretti and R. Nevzorov,
 Phys. Lett. {\bf B650}, 57-64 (2007),  hep-ph/0701064.}

\nref\HenWyl{M. Henningson and N. Wyllard,
   JHEP 0707:084, (2007),  arXiv:0706.2803 [hep-th].}

\nref\GuLuMi{S. Gurrieri, A. Lukas and A. Micu, JHEP 0712:081
(2007), arXiv:0709.1932.}

\nref\BCCS{F. Bernardoni, S.L. Cacciatori, Bianca L. Cerchiai and A.
Scotti, J. Math. Phys. {\bf 49}, 012107 (2008), arXiv:0710.0356.}

\nref\Kallosh{R. Kallosh and M. Soroush,
Nucl. Phys. {\bf B801}, 25-44 (2008), arXiv:0802.4106;\nl R. Kallosh
and T. Kugo, arXiv:0811.3414.}

\nref\NuFuPe{J.F. Nunez, W.G. Fuertes and  A.M. Perelomov,
Quantum trigonometric Calogero-Sutherland model and irreducible characters for the exceptional algebra E(8),
  arXiv:0806.1011 [math-ph].}

\nref\Sati{H. Sati, OP**2 bundles in M-theory, arXiv:0807.4899.}

\nref\Mizo{Sh. Mizoguchi,
 Localized Modes in Type II and Heterotic Singular Calabi-Yau Conformal Field Theories,
 arXiv:0808.2857.}

\nref\HowKin{R. Howl and S.F. King, JHEP,  0801:060 (2008).}

\nref\DufSam{A. Le Diffon and H. Samtleben,
Supergravities without an Action: Gauging the Trombone,
arXiv:0809.5180.}

\lref\Dobesix{V.K. Dobrev, Invariant Differential Operators for Non-Compact
Lie Groups: the $E_{6(-14)}$ case, arXiv:0812.2655 [math-ph],
Invited talks at V School in Modern Mathematical Physics, Belgrade, Serbia,
6-17.7.2008, to appear in the Proceedings.}

\lref\Helg{S. Helgason, {\it Differential Geometry, Lie Groups and
Symmetric Spaces},\hfill\break (Academic Press, New York, 1978).}

\lref\Dobmul{V.K. Dobrev, Lett. Math. Phys. {\bf 9}, 205-211
(1985).}

\lref\PeSo{V.B. Petkova and G.M. Sotkov, Lett. Math. Phys.
{\bf 8} (1984) 217-226; Erratum-ibid. {\bf 9} (1985) 83.}

\lref\Gilm{R. Gilmore,  {\it  Lie groups, Lie algebras, and some of
their applications}, (New York,  Wiley, 1974).}

\lref\Dobads{V.K. Dobrev,
``Intertwining operator realization of the AdS/CFT correspondence'',
Nucl. Phys. {\bf B553} (1999) 559-582; hep-th/9812194.}

\lref\TMP{I.T. Todorov, M.C. Minchev and V.B. Petkova,
{\it Conformal invariance in QFT}, Scuola Normale Superiore (Pisa, 1978).}

\lref\Dobc{V.K. Dobrev,
J. Math. Phys. {\bf 26} (1985) 235-251; first as
ICTP Trieste preprint IC/83/36 (1983).}

 \lref\Mack{G. Mack, Commun. Math. Phys. {\bf 55} (1977) 1-28.}

\lref\GuKoNi{M. Gunaydin, K. Koepsell and H. Nicolai, Commun. Math. Phys. {\bf 221},
57-76 (2001) hep-th/0008063.}

\lref\FerGun{S. Ferrara and M. Gunaydin, Int. J. Mod. Phys. {\bf A13}, 2075-2088 (1998).}

\lref\Guna{M. Gunaydin, Mod. Phys. Lett. {\bf A8} (1993) 1407-1416.}

\lref\Mackder{G. Mack and M. de Riese, J. Math. Phys. {\bf 48} (2007) 052304;
hep-th/0410277v2.}

\lref\HC{Harish-Chandra, "Representations of semisimple Lie groups:
IV,V", Am. J. Math. {\bf 77} (1955) 743-777, {\bf 78} (1956) 1-41.}

\lref\Har{Harish-Chandra, ``Discrete series for semisimple Lie
groups: II'', Ann. Math. {\bf 116} (1966) 1-111.}

\lref\KnSt{A.W. Knapp and E.M. Stein, ``Intertwining operators for
semisimple groups'', Ann. Math. {\bf 93} (1971) 489-578; II : Inv.
Math. {\bf 60} (1980) 9-84.}

\lref\BGG{I.N. Bernstein, I.M. Gel'fand and S.I. Gel'fand,
 Funkts. Anal. Prilozh. {\bf 5} (1) (1971) 1-9; English
translation: Funct. Anal. Appl. {\bf 5} (1971) 1-8.}

\lref\War{G. Warner, {\it Harmonic Analysis on Semi-Simple Lie
Groups I}, (Springer, Berlin, 1972).}

\lref\Lan{R.P. Langlands, {\it On the classification of irreducible
representations of real algebraic groups}, Math. Surveys and
Monographs, Vol.  31 (AMS, 1988), first as IAS Princeton preprint
(1973).}

\lref\Zhea{D.P. Zhelobenko, {\it Harmonic Analysis on Semisimple
Complex Lie Groups}, (Moscow, Nauka, 1974, in Russian).}

\lref\Dix{J. Dixmier, {\it Enveloping Algebras}, (North Holland, New
York, 1977).}

\lref\Bourb{N. Bourbaki, {\it Groupes at alg\`{e}bres de Lie,
Chapitres 4,5 et 6}, (Hermann, Paris, 1968).}

\lref\KnZu{A.W. Knapp and G.J. Zuckerman, ``Classification theorems
for representations of semisimple groups'',
 in: Lecture Notes in Math., Vol. 587 (Springer, Berlin,
1977) pp. 138-159; ~``Classification of irreducible tempered
representations of semisimple groups'', Ann. Math. {\bf 116} (1982)
389-501.}

\lref\DMPPT{V.K. Dobrev, G. Mack, V.B. Petkova, S.G. Petrova and
I.T. Todorov, {\it Harmonic Analysis on the  $n$-Dimensional Lorentz
Group and Its Applications to Conformal Quantum Field Theory},
Lecture Notes in Physics, Vol. 63  (Springer-Verlag,
 Berlin-Heidelberg-New York, 1977);
V.K. Dobrev and V.B. Petkova, Rept. Math. Phys. {\bf 13}, 233-277
(1978).}

\lref\Knapp{A.W. Knapp, {\it Representation Theory of Semisimple
Groups (An Overview Based on Examples)}, (Princeton Univ. Press,
1986).}

\lref\Dob{V.K. Dobrev,
Rept. Math. Phys. {\bf 25}, 159-181 (1988) ; first as ICTP Trieste
preprint IC/86/393 (1986).}

\lref\DobPet{V.K. Dobrev and V.B. Petkova, Phys. Lett. {\bf B162},
127-132 (1985); Lett. Math. Phys. {\bf 9}, 287-298 (1985); Fortsch.
Phys. {\bf 35}, 537-572 (1987).}

\lref\Dobsusy{V.K. Dobrev, Phys. Lett. {\bf B186}, 43-51 (1987);
J. Phys. {\bf A35}, 7079-7100 (2002), hep-th/0201076; Phys. Part.
Nucl. {\bf 38}, 564-609 (2007),
  hep-th/0406154;
V.K. Dobrev and A.Ch. Ganchev, Mod. Phys. Lett. {\bf A3}, 127-137
(1988).}

\lref\Dobso{V.K. Dobrev,
J. Phys. {\bf A39} (2006) 5995-6020; hep-th/0512354.}

\lref\Dobsin{V.K. Dobrev,
Lett. Math. Phys. {\bf 22}, 251-266 (1991).}

\lref\Dobqg{V.K. Dobrev,
J. Math. Phys. {\bf 33}, 3419-3430 (1992) 
J. Phys.  {\bf A26}, 1317-1334 (1993), first as G\"{o}ttingen
University preprint, (July 1991);
~
J. Phys.  {\bf A27}, 4841-4857 (1994), Erratum-ibid.  {\bf A27},
6633-6634 (1994), hep-th/9405150;
~ 
Phys. Lett. {\bf B341}, 133-138 (1994) Erratum-ibid. {\bf B346}, 427
(1995);
~V.K. Dobrev and P.J. Moylan, 
Phys. Lett. {\bf B315}, 292-298 (1993) .}  

\lref\Sata{I. Satake,
Ann. Math. {\bf 71} (1960) 77-110.}

\lref\Dobp{V.K. Dobrev, in preparation.}

\lref\Dobinv{V.K. Dobrev, 
Rev. Math. Phys. {\bf 20} (2008) 407-449; hep-th/0702152; ICTP
Trieste preprint IC/2007/015.}

\lref\Dobpeds{V.K. Dobrev, 
J. Phys. A: Math. Theor. {\bf 41} (2008) 425206; arXiv:0712.4375
[hep-th].}


\lref\EHW{T. Enright, R. Howe and W. Wallach, "A classification of
unitary highest weight modules", in: {\it Representations of
Reductive Groups}, ed. P. Trombi (Birkh\"auser, Boston, 1983) pp.
97-143.}

\lref\Witten{E. Witten, "Conformal Field Theory in Four and Six
Dimensions", arXiv:0712.0157.}



\vskip 1.5cm

\centerline{{\tfont   Exceptional   Lie Algebra E$_{\hbox{\bf 7(-25)}}$ }}
\vskip 2truemm
\centerline{{\sfont (Multiplets and Invariant Differential Operators)}}

\vskip 1.5cm

\centerline{{\bf V.K. Dobrev}} \vskip 0.5cm

\centerline{Institute for Nuclear Research and Nuclear Energy}
\centerline{Bulgarian Academy of Sciences} \centerline{72
Tsarigradsko Chaussee, 1784 Sofia, Bulgaria} \centerline{(permanent
address)}

\vskip 0.5cm \centerline{and}

\vskip 0.5cm

 \centerline{The Abdus Salam International Centre for
Theoretical Physics} \centerline{P.O. Box 586, Strada Costiera 11}
\centerline{34014 Trieste, Italy}

\vskip 1.5cm

 \centerline{{\bf Abstract}}
\midinsert\narrower{\male In the present paper we continue the
project of systematic   construction of invariant
differential operators on the example of the non-compact
exceptional algebra  $E_{7(-25)}\,$. Our choice of this particular algebra is motivated by the fact that it
belongs to a narrow class of algebras, which we call 'conformal Lie algebras', which have
very similar properties to the conformal algebras of $n$-dimensional Minkowski space-time.
This class of algebras is identified and summarized in a table.
Another motivation is related to the AdS/CFT correspondence.\nl
We give the multiplets of indecomposable elementary representations, including the necessary data
for all relevant invariant differential operators.
 }\endinsert

\vskip 1.5cm

\newsec{Introduction}

\subsec{Generalities}

\nt
Recently, there was more interest in the study and applications of
exceptional Lie groups, cf., e.g., \refs{\KanLan-\DufSam}.\nl
Thus, in the development of our project \Dobinv{} of systematic   construction of invariant
differential operators for non-compact Lie groups we decided to give priority to some
exceptional Lie groups. We start with the more interesting ones -  the
only two exceptional Lie groups/algebras that have highest/lowest weight representations,
namely, ~$E_{6(-14)}\,$, cf. \Dobesix, and ~$E_{7(-25)}\,$, which we consider
in the present paper.

In fact, there are additional motivations for the choice of ~$E_{7(-25)}\,$,
namely, it belongs to a narrow class of algebras, which we call 'conformal Lie algebras', which have
very similar properties to the conformal algebras ~$so(n,2)$~ of $n$-dimensional Minkowski space time.
Another motivation is related to the AdS/CFT correspondence.

Thus, we expand our motivations in the next Subsection, where we also give the table
of the conformal Lie algebras.

Further the paper is organized as follows. In section 2 we give the
preliminaries, actually recalling and adapting facts from \Dobinv.
In Section 3 we specialize to the ~$E_{7(-25)}\,$~ case. In Section
4 we present our results on the multiplet classification of the
representations and \idos\ between them. In Subsection 4.1 we make a brief
interpretation of our results to relate to the usual conformal algebras.

\subsec{Motivation: the class of conformal Lie algebras}

\nt
The group-theoretical interpretation of the AdS/CFT correspondence \Dobads{},
or more general ~{\it holography},
involves two standard decompositions valid for any non-compact
semi-simple Lie group ~$G$~ or Lie algebra ~$\cg$ (also
super-group/algebra)~:~ the ~{\it Iwasawa} decomposition:
\eqn\iwas{ G ~=~ K\,A\,N \ , \quad \cg ~=~ \ck\oplus\ca\oplus\cn \ ,}
where ~$K$~ is the maximal compact subgroup of ~$G$,
~$A$~ is abelian simply connected subgroup of ~$G$,\foot{Actually, ~$A ~\cong~ SO(1,1) \times \cdots \times SO(1,1)$, ~$r = \dim\,A$~ copies.}
 ~$N$~ is a nilpotent simply connected
subgroup of ~$G$~ preserved by the action of ~$A$, (and similarly for the algebra
decomposition),\foot{The group decomposition is global which means that each element ~$g$~ of ~$G$~ can be represented
by the group multiplication of three elements from the respective subgroups ~$g = kan$, ~$k\in K$,
~$a\in A$, ~$n\in N$. Similarly, each element ~$W\in \cg$, can be represented as the   sum
~$W = X \oplus Y \oplus Z$, ~$X\in\ck$, ~$Y\in\ca$, ~$Z\in\cn$.}
~~and the ~{\it Bruhat} decomposition:
\eqn\bruhm{ G ~=~ M\,A\,N\,\tN \ , \quad \cg ~=~ \cm\oplus\ca\oplus\cn \oplus\tcn\ ,}
where ~$M$~ is a maximal subgroup of ~$K$~ that commutes with ~$A$, ~$\tN$~ is a subgroup conjugate to $N$ by the Cartan
involution.\foot{This group decomposition is almost global, which means that the decomposition ~$g = man\tn$, ($m\in M$, $\tn\in\tN$)
is valid except for a subset of $G$ of lower dimensionality. But the  algebra decomposition ~~$W = U \oplus Y \oplus Z \oplus \tZ$,
($U\in\cm$,  ~$\tZ\in\tcn$), ~ is valid as above for each element ~$W\in \cg$.}
The Iwasawa decomposition is used to define induced representations on the bulk, which in this approach is represented by the
solvable subgroup ~$AN$, while the Bruhat decomposition is used to define induced representations on the conformal boundary,
i.e., on space-time, represented by the subgroup ~$N$, \Dobads.

The application of the group-theoretical approach in \Dobads{} for
the Euclidean conformal group ~$G= SO(n+1,1)$~ was facilitated  by
 the fact that in the group-subgroup chain ~$G\supset K \supset M$~ the
subgroups were sufficiently large: ~$K=SO(n+1)$, ~$M=SO(n)$. Thus,
there was not much freedom when embedding representations, in
particular, embedding the representations of ~$SO(n)$~ into those of
~$SO(n+1)$.

Since the non-compact exceptional Lie algebra ~$E_{7(+7)}\,$~ was prominently used recently, cf. \Kallosh,
we would like to apply similar interpretation to its holography. However, there is the  problem
of subgroups being not large enough. In fact, while the maximal compact subalgebra is
~$\ck = su(8)$, the corresponding subalgebra ~$\cm$~ is null, ~$\cm = \{0\}$,  and the Bruhat decomposition
is just ~$\cg ~=~ \ca\oplus\cn \oplus\tcn$. The reason
is that ~$E_{7(+7)}\,$~ is maximally split, in fact, it is just the restriction to the real numbers
of the complex Lie algebra ~$E_7\,$.

In fact, that would be a general problem in the case when the dimension ~$r$~ of the subalgebra ~$\ca$, called ~{\it real rank} or
~{\it split rank}, is bigger than 1. But that also contains possible solutions of the problem, since when ~$r>1$~
the algebra under consideration has more Bruhat decompositions,
in fact, their number is ~$2^r-1$. They are written in a similar way (writing only the algebra version):
\eqn\bruhma{ \cg ~=~ \cm'\oplus\ca'\oplus\cn' \oplus\tcn'\ ,}
so that ~$\cm' \supset \cm$, ~$\ca' \subset \ca$, ~$\cn' \subset \cn$, ~$\tcn' \subset \tcn$.
Especially useful are the so-called 'maximal' decompositions, when ~$\dim\ca'=1$, since they represent closer the case $r=1$,
and the idea that the dimensions of the bulk (with Lie algebra ~$\ca'\cn'$)
and the boundary (with Lie algebra ~$\cn'$) should differ by 1.

In the case of ~$E_{7(+7)}\,$~ there are several suitable Bruhat decompositions \Dobinv{}\foot{The number of maximal Bruhat
decompositions is equal to ~$r\,$.}~:
\eqn\bruhs{ E_{7(+7)} ~=~ \cm_1\oplus\ca_1\oplus\cn_1 \oplus\tcn_1 \ , \qquad
\cm_1 ~=~ so(6,6), ~~\dim\ca_1=1, ~~\dim \cn_1 = \dim\tcn_1 = 33 }
\eqn\bruhz{ E_{7(+7)} ~=~ \cm_2\oplus\ca_2\oplus\cn_2 \oplus\tcn_2 \ , \qquad
\cm_2 ~=~ E_{6(+6)}\ , ~~\dim\ca_2=1, ~~\dim \cn_2 = \dim\tcn_2 = 27 }

Due to the presence of the subalgebra ~$so(6,6)$~ the first case deserves separate study.
The decomposition \bruhz{} is mentioned, though not in our context,
in \GuKoNi, where it is called three-graded decomposition,
and in \Kallosh, thus, it may be useful  in applications to supergravity.
However, instead of using the Bruhat decomposition \bruhz{},   we shall use
another non-compact real form of ~$E_7\,$, namely,  the Lie algebra ~$E_{7(-25)}\,$.

There are several motivations to use the non-compact exceptional Lie algebra ~$E_{7(-25)}\,$.
Unlike ~$E_{7(+7)}\,$~ it has discrete series representations. Even more important is that it
is one of two exceptional non-compact groups that have highest/lowest weight representations.\foot{The other
one is ~$E_{6(-14)}\,$ which we have also started to study, \Dobesix{}.}

The groups that have highest/lowest weight representations are called Hermitian symmetric spaces \HC.
The corresponding non-compact Lie algebras are:
\eqn\hermsym{
su(m,n), ~~so(n,2), ~~sp(2n,R), ~~so^*(2n), ~~E_{6(-14)}\,, ~~E_{7(-25)} \ , }
cf., e.g., \Helg. The practical criterion is that in these cases, the
maximal compact subalgebras are of the form:
\eqn\maxc{\ck ~=~ \ck' \oplus so(2)\ }
The most widely used of these algebras are the conformal algebras ~$so(n,2)$~ in $n$-dimensional Minkowski space-time.
In that case, there is a maximal Bruhat decomposition that has direct physical meaning:
\eqn\bruhc{\eqalign{ so(n,2) ~~=&~ \cm_c\,\oplus\,\ca_c\,\oplus\,\cn_c\, \oplus\,\tcn_c \ , \cr
&\cm_c ~=~ so(n-1,1) \ , ~~\dim\ca_c=1, ~~\dim \cn_c = \dim\tcn_c = n }}
Indeed, ~$\cm_c ~=~ so(n-1,1)$~ is the Lorentz algebra of $n$-dimensional Minkowski space-time, the subalgebra
~$\ca_c ~=~ so(1,1)$~ represents the dilatations, the conjugated subalgebras ~$\cn_c\,$, $\tcn_c\,$~
are the algebras of translations, and special conformal transformations, both being isomorphic to
$n$-dimensional Minkowski space-time.\foot{The Bruhat-decomposition interpretation of the conformal subgroups/subalgebras
was done first in the Euclidean case, cf. \DMPPT, then in the Minkowski case, cf. \Dobc,
for the general picture see  \Dob.}

There are other special features which are important. In particular, the complexification of the maximal compact subgroup
coincides with the complexification of the first two factors of the Bruhat decomposition \bruhc:
\eqn\compl{\ck^\bac ~=~ so(n,\bbc) \oplus so(2,\bbc) ~=~ so(n-1,1)^\bac \oplus so(1,1)^\bac = \cm_c^\bac \oplus\ca_c^\bac }
In particular, the coincidence of the complexification of the semi-simple subalgebras in \compl{}, ~$so(n,\bbc) ~=~ so(n-1,1)^\bac$,
means that the sets of finite-dimensional (nonunitary) representations of
~$\cm_c$~ are in 1-to-1 correspondence with the finite-dimensional
(unitary) representations of ~$so(n)$. The latter leads to the fact that the induced representations that we consider
in this paper (and which are of the type that is mostly used in physics), cf. next Section,
are representations of finite $\ck$-type \HC.\nl
The role of the abelian factors in \compl{} for the construction of highest/lowest weight representations was singled out
first in \Mack.

It turns out that some of the algebras in \hermsym{} share the  above-mentioned special properties of ~$so(n,2)$.
That is why, in view of applications to physics, these algebras, together with the appropriate Bruhat decompositions
should be called 'conformal Lie algebras', (resp. 'conformal Lie groups' in the group setting). We display all these algebras in the following table:

\bigskip
\vbox{\offinterlineskip
{\bf Table of conformal Lie algebras}\medskip
\halign{\baselineskip12pt
\strut\vrule#\hskip0.1truecm &
#\hfil&
\vrule#\hskip0.1truecm &
#\hfil&
\vrule#\hskip0.1truecm  &
#\hfil&
\vrule#\hskip0.1truecm  &
#\hfil&
\hskip0.1truecm  \vrule#\cr
\tablerule
&&&&&&&&\cr
&~ $\cg$
&&~ $\ck'$
&&~  $\cm_c $
&&~$\dim_\bbr\,\cn_c$&\cr
&&&&& &&&\cr
\tablerule
&           &&&&&&&\cr
&~$su(n,n)$ && ~$su(n)\oplus su(n)$ && ~$sl(n,\bbc)_\bbr$
&& ~$n^2$ &\cr
&&&&&&&&\cr
\tablerule
&           &&&&&&&\cr
&~$so(n,2)$ && ~$so(n)$ && ~$so(n-1,1)$ && ~$n$ &\cr
&~$n>4$ &&&&&&&\cr
&&&&&&&&\cr
\tablerule
&           &&&&&&&\cr
&~$sp(n,\bbr)$ && ~$su(n)$ && ~$sl(n,\bbr)$ && ~$\ha (n+1)n$ &\cr
&~$n\geq 2$ &&&&&&&\cr
&&&&&&&&\cr
\tablerule
&           &&&&&&&\cr
&~$so^*(2n)$   && ~$su(n)$ && ~$su^*(n)$ && ~$\ha n(n-1)$ &\cr
&~$n$~-~even, ~$n\geq 6$ &&&&&&&\cr
&&&&&&&&\cr
\tablerule
&           &&&&&&&\cr
&~$E_{7(-25)}$   && ~$e_6$ && ~$E_{6(-26)}$ && ~$27$ &\cr
&&&&&&&&\cr
\tablerule
}}
\nt
where we display only the semisimple part ~$\ck'$~ of ~$\ck$, ~$sl(n,\bbc)_\bbr$~ denotes $sl(n,\bbc)$ as a real Lie algebra, (thus,
~$(sl(n,\bbc)_\bbr)^\bac = sl(n,\bbc)\oplus sl(n,\bbc)$), ~$e_6$~ denotes the compact real form of ~$E_6\,$, and we have imposed
restrictions to avoid coincidences or inconsistency due to well known isomorphisms:
~$so(1,2) \cong sp(1,\bbr) \cong su(1,1) $, ~~$so(2,2) \cong so(1,2)\oplus so(1,2)$,
~$so(3,2)\cong sp(2,\bbr)$, ~$so(4,2)\cong su(2,2)$, ~$so^*(4)\cong so(3) \oplus so(2,1)$,
~$so^*(8) \cong so(6,2)$.

The same class was identified from different considerations in \Guna{}, where these groups/algebras
 were called 'conformal groups  of simple Jordan algebras'. It was identified
 from still different considerations also in \Mackder, where
the objects of the class were called simple space-time symmetries generalizing conformal symmetry.

 Finally, we should mention that the algebra ~$E_{7(-25)}\,$~ was applied to the classification of
 orbits of BPS black holes in N=2 Maxwell-Einstein supergravity theories \FerGun.

With these motivations in mind we continue with the algebra ~$E_{7(-25)}\,$~
with the following maximal Bruhat decomposition:
\eqn\bruhzz{ E_{7(-25)} ~=~ \cm'\oplus\ca'\oplus\cn' \oplus\tcn' \ , \qquad
\cm' ~=~ E_{6(-26)}\ , ~~\dim\ca'=1, ~~\dim \cn' = \dim\tcn' = 27 }
The careful reader may notice that the above Bruhat decomposition
is a Wick-rotation of the corresponding one for ~$E_{7(+7)}\,$, \bruhz{}, yet there are
crucial differences in their properties.

The next Section contains preliminaries which are general for our programme started in \Dobinv.

\newsec{Preliminaries}

\nt This Section can be read independently from the Introduction.
Let $G$ be a  semisimple non-compact Lie group, and  $K$ a
maximal compact subgroup of $G$. Then we have an Iwasawa
decomposition ~$G=KAN$, where ~$A$~ is abelian simply
connected vector subgroup of ~$G$, ~$N$~ is a nilpotent simply
connected subgroup of ~$G$~ preserved by the action of ~$A$.
Further, let $M$ be the centralizer of $A$ in $K$. Then the
subgroup ~$P_0 ~=~ M A N$~ is a minimal parabolic subgroup of
$G$. A parabolic subgroup ~$P ~=~ M' A' N'$~ is any subgroup of $G$
(including $G$ itself) which contains a minimal parabolic
subgroup.\foot{The number of non-conjugate parabolic subgroups is
~$2^r$, where $r=\rank\,A$, cf., e.g., \War.}

The importance of the parabolic subgroups comes from the fact that
the representations induced from them generate all (admissible)
irreducible representations of $G$ \Lan. For the classification of
all irreducible representations it is enough to use only the
so-called {\it cuspidal} parabolic subgroups ~$P=M'A'N'$, singled
out by the condition that ~rank$\, M' =$ rank$\, M'\cap K$
\Zhea,\KnZu, so that $M'$ has discrete series representations \Har.
However, often induction from non-cuspidal parabolics is also
convenient, cf. \EHW,\Dobinv,\Dobpeds.

Let ~$\nu$~ be a (non-unitary) character of ~$A'$, ~$\nu\in\ca'^*$,
let ~$\mu$~ fix an irreducible representation ~$D^\mu$~ of ~$M'$~ on
a vector space ~$V_\mu\,$.

  We call the induced
representation ~$\chi =$ Ind$^G_{P}(\mu\otimes\nu \otimes 1)$~ an
~{\it elementary representation} of $G$ \DMPPT. (These are called
{\it generalized principal series representations} (or {\it limits
thereof}) in \Knapp.) Their spaces of functions are: \eqn\fun{
\cc_\chi ~=~ \{ \cf \in C^\infty(G,V_\mu) ~ \vr ~ \cf (gman) ~=~
e^{-\nu(H)} \cdot D^\mu(m^{-1})\, \cf (g) \} } where ~$a= \exp(H)\in
A'$, ~$H\in\ca'\,$, ~$m\in M'$, ~$n\in N'$. The representation
action is the $left$ regular action: \eqn\lrr{ (\ct^\chi(g)\cf) (g')
~=~ \cf (g^{-1}g') ~, \quad g,g'\in G\ .}

For our purposes we need to restrict to ~{\it maximal}~ parabolic
subgroups ~$P$, (so that $\rank\,A'=1$), that may not be cuspidal.
For the representations that we consider the character ~$\nu$~ is
parameterized by a real number ~$d$, called the conformal weight or
energy.

Further, let ~$\mu$~ fix a discrete series representation ~$D^\mu$~
of $M'$ on the Hilbert space ~$V_\mu\,$, or the so-called limit of a
discrete series representation (cf. \Knapp). Actually, instead of
the discrete series we can use the finite-dimensional (non-unitary)
representation of $M'$ with the same Casimirs.

An important ingredient in our considerations are the ~{\it
highest/lowest weight representations}~ of ~$\cg$. These can be
realized as (factor-modules of) Verma modules ~$V^\L$~ over
~$\cg^\bac$, where ~$\L\in (\ch^\bac)^*$, ~$\ch^\bac$ is a Cartan
subalgebra of ~$\cg^\bac$,  weight ~$\L = \L(\chi)$~ is determined
uniquely from $\chi$ \Dob. In this setting we can consider also
unitarity, which here means positivity w.r.t. the Shapovalov form in
which the conjugation is the one singling out $\cg$ from $\cg^\bac$.

Actually, since our ERs may be induced from finite-dimensional
representations of ~$\cm'$~ (or their limits) the Verma modules are
always reducible. Thus, it is more convenient to use ~{\it
generalized Verma modules} ~$\tV^\L$~ such that the role of the
highest/lowest weight vector $v_0$ is taken by the
(finite-dimensional) space ~$V_\mu\,v_0\,$. For the generalized
Verma modules (GVMs) the reducibility is controlled only by the
value of the conformal weight $d$. Relatedly, for the \idos{} only
the reducibility w.r.t. non-compact roots is essential.

One main ingredient of our approach is as follows.  We group the
(reducible) ERs with the same Casimirs  in sets called ~{\it
multiplets} \Dobmul,\Dob{}. The multiplet corresponding to fixed
values of the Casimirs may be depicted as a connected graph, the
vertices of which correspond to the reducible ERs and the lines
between the vertices correspond to intertwining operators.\foot{For
simplicity only the operators which are not compositions of other
operators are depicted.} The explicit parametrization of the
multiplets and of their ERs is important for understanding of the
situation.

In fact, the multiplets contain explicitly all the data necessary to
construct the \idos{}. Actually, the data for each \ido{} consists
of the pair ~$(\b,m)$, where $\b$ is a (non-compact) positive root
of ~$\cg^\bac$, ~$m\in\bbn$, such that the BGG \BGG{} Verma module
reducibility condition (for highest weight modules) is fulfilled:
\eqn\bggr{    (\L+\r, \b^\vee ) ~=~ m \ , \quad \b^\vee \equiv 2 \b
/(\b,\b) \ .} When \bggr{} holds then the Verma module with shifted
weight ~$V^{\L-m\b}$ (or ~$\tV^{\L-m\b}$ ~ for GVM and $\b$
non-compact) is embedded in the Verma module ~$V^{\L}$ (or
~$\tV^{\L}$). This embedding is realized by a singular vector
~$v_s$~ determined by a polynomial ~$\cp_{m,\b}(\cg^-)$~ in the
universal enveloping algebra ~$(U(\cg_-))\ v_0\,$, ~$\cg^-$~ is the
subalgebra  of ~$\cg^\bac$  generated by the negative root
generators \Dix.
 More explicitly, \Dob, ~$v^s_{m,\b} = \cp_{m,\b}\, v_0$ (or ~$v^s_{m,\b} = \cp_{m,\b}\, V_\mu\,v_0$ for GVMs).\foot{For
explicit expressions for singular vectors we refer to \Dobsin.} Then
there exists \Dob{} an \ido{} \eqn\lido{\cd_{m,\b} ~:~
\cc_{\chi(\L)} ~\llr ~  \cc_{\chi(\L-m\b)} } given explicitly by:
\eqn\mido{\cd_{m,\b} ~=~ \cp_{m,\b}(\widehat{\cg^-})  } where
~$\widehat{\cg^-}$~ denotes   the $right$ action on the functions
~$\cf$, cf. \fun.

\newsec{The non-compact Lie algebra $E_{7(-25)}$}

\nt Let ~$\cg=E_{7(-25)}$. The maximal compact subgroup is ~$\ck
\cong e_6\oplus so(2)$, ~$\dim_\bbr\,\cp = 54$, ~$\dim_\bbr\,\cn
= 51$. This real form has discrete series representations and
highest/lowest weight representations.

The split rank is equal to 3, while ~$\cm \cong so(8)$.

The Satake diagram is \Sata:
\eqn\satsevb{ \downcirc{{\a_1}}
\riga\black{\a_3} \riga
\black{{\a_4}}\kern-8pt\raise11pt\hbox{$\vert$}
\kern-3.5pt\raise22pt\hbox{$\bullet {{\scriptstyle{\a_2}}}$}
\riga\black{{\a_5}} \riga\downcirc{{\a_6}} \riga\downcirc{{\a_7}}}
Thus, the reduced root system is presented by a Dynkin-Satake
diagram looking like the ~$C_3$~ Dynkin diagram: \eqn\satsevc{
\downcirc{{\l_1}} \Longrightarrow\downcirc{{\l_2}}
\riga\downcirc{{\l_3}}} but the short roots have multiplicity $8$
(the long - multiplicity $1$). Going to the $C_3$ diagram we drop
the black nodes, (they give rise to $\cm$), while
$\a_1\,,\a_6\,,\a_7\,$, are mapped to $\l_1\,,\l_2\,,\l_3\,$, resp.,
of \satsevc.

We choose a ~{\it maximal} parabolic ~$\cp=\cm'\ca'\cn'$~ such that
~$\ca'\cong so(1,1)$, while the factor ~$\cm'$~ has the same
finite-dimensional (nonunitary) representations as the
finite-dimensional (unitary) representations of  the semi-simple
subalgebra of   ~$\ck$, i.e., ~$\cm' = E_{6(-26)}\,$, cf. \Dobinv.
Thus, these induced representations are representations of finite
$\ck$-type \HC. Relatedly, the number of ERs in the corresponding
multiplets is equal to ~$\vr W(\cg^\bac,\ch^\bac)\vr\, /\, \vr
W(\ck^\bac,\ch^\bac)\vr ~=~ 56$, cf. \Bourb, where ~$\ch$~ is a
Cartan subalgebra of both ~$\cg$~ and ~$\ck$. Note also that
~$\ck^\bac \cong\cm'^\bac \oplus \ca'^\bac$. Finally, note that
~$\dim_\bbr\,\cn' = 27$.

We label   the signature of the ERs of $\cg$   as follows:
\eqn\sgnd{  \chi ~=~ \{\, n_1\,, \ldots,\, n_{6}\,;\, c\, \} \ ,
\qquad n_j \in \bbn\ , \quad c = d- 9 } where the last entry of ~$\chi$~ labels
the characters of $\ca'\,$, and the first $6$ entries are labels of
the finite-dimensional nonunitary irreps of $\cm'\,$, (or of the
finite-dimensional unitary irreps of the ~$e_6$).

 The reason to use the parameter ~$c$~ instead of ~$d$~ is that the
parametrization of the ERs in the multiplets is given in a simpler way,
as we shall see.

Further, we need the root system of the complex algebra ~$E_7\,$.
With Dynkin diagram enumerating the simple roots ~$\a_i$~ as in
\satsevb, the positive roots are:\nl first there are ~21~
roots~forming~the~positive~roots~of ~$sl(7)$~ with~simple~roots~~
$\a_1,\a_3,\a_4,\a_5$, $\a_6,\a_7\,$, ~~then ~21~ roots which are roots
of the ~$E_6$~ subalgebra and include the non-$sl(7)$ root ~$\a_2$~:
\eqna\satsevz
$$\eqalignno{
&\a_2\ , ~~\a_2+\a_4\ , ~~\a_2+\a_4+\a_3\ , ~~\a_2+\a_4+\a_5\ ,
~~\a_2+\a_4+\a_3+\a_5\ , &\satsevz {}\cr &\a_2+\a_4+\a_3+\a_1\ ,
~~\a_2+\a_4+\a_5+\a_6\ , ~~\a_2+\a_4+\a_3+\a_5+\a_1\ , \cr&
\a_2+\a_4+\a_3+\a_5+\a_6\ , ~~\a_2+\a_4+\a_3+\a_5+\a_1+\a_6 \ , ~~~~
\a_2+2\a_4+\a_3+\a_5 \ , \cr &\a_2+2\a_4+\a_3+\a_5+\a_1\ ,
~~\a_2+2\a_4+\a_3+\a_5+\a_6\ , ~~\a_2+2\a_4+\a_3+\a_5+\a_1+\a_6\ ,
\cr & \a_2+2\a_4+2\a_3+\a_5+\a_1\ , ~~\a_2+2\a_4+\a_3+2\a_5+\a_6\ ,
~~\a_2+2\a_4+2\a_3+\a_5+\a_1+\a_6\ ,\cr&
\a_2+2\a_4+\a_3+2\a_5+\a_1+\a_6\ ,
~~\a_2+2\a_4+2\a_3+2\a_5+\a_1+\a_6\ ,\cr&
\a_2+3\a_4+2\a_3+2\a_5+\a_1+\a_6\ , ~~
2\a_2+3\a_4+2\a_3+2\a_5+\a_1+\a_6  \ , }$$
finally there are the following ~21~ roots including the non-$E_6$ root ~$\a_7$~:
\eqna\satsevy
$$\eqalignno{&\a_2+\a_4+\a_5+\a_6+\a_7\ ,
~~  \a_2+\a_4+\a_3+\a_5+\a_6+\a_7\ , &\satsevy {}\cr&
\a_2+\a_4+\a_3+\a_5+\a_1+\a_6+\a_7 \ , \cr&
   \a_2+2\a_4+\a_3+\a_5+\a_6+\a_7\ ,
~~\a_2+2\a_4+\a_3+\a_5+\a_1+\a_6+\a_7\ , \cr &
   \a_2+2\a_4+\a_3+2\a_5+\a_6+\a_7\ ,
~~\a_2+2\a_4+2\a_3+\a_5+\a_1+\a_6+\a_7\ ,\cr&
\a_2+2\a_4+\a_3+2\a_5+\a_1+\a_6+\a_7\
,~~\a_2+2\a_4+2\a_3+2\a_5+\a_1+\a_6+\a_7\ ,\cr&
\a_2+3\a_4+2\a_3+2\a_5+\a_1+\a_6+\a_7\ , ~~
2\a_2+3\a_4+2\a_3+2\a_5+\a_1+\a_6+\a_7\ ,\cr &
   \a_2+2\a_4+\a_3+2\a_5+2\a_6+\a_7\ ,\cr&
\a_2+2\a_4+\a_3+2\a_5+\a_1+2\a_6+\a_7\
,~~\a_2+2\a_4+2\a_3+2\a_5+\a_1+2\a_6+\a_7\ ,\cr&
\a_2+3\a_4+2\a_3+2\a_5+\a_1+2\a_6+\a_7\ , ~~
2\a_2+3\a_4+2\a_3+2\a_5+\a_1+2\a_6+\a_7\ ,\cr&
\a_2+3\a_4+2\a_3+3\a_5+\a_1+2\a_6+\a_7\ , ~~
2\a_2+3\a_4+2\a_3+3\a_5+\a_1+2\a_6+\a_7\ ,\cr&
    2\a_2+4\a_4+2\a_3+3\a_5+\a_1+2\a_6+\a_7\ ,\cr&
   2\a_2+4\a_4+3\a_3+3\a_5+\a_1+2\a_6+\a_7\ ,\cr&
    2\a_2+4\a_4+3\a_3+3\a_5+2\a_1+2\a_6+\a_7 ~=~ \tilde{\a}\ ,
}$$
where ~$\ta$~ is the highest root of the $E_7$ root system.

 The differential
intertwining operators that give the multiplets correspond to  the noncompact
roots, and since we shall use the latter extensively, we introduce more
compact notation for them. Namely, the nonsimple roots  will be
denoted in a self-explanatory way as follows:
\eqna\nota
$$\eqalignno{
&\a_{ij} ~=~ \a_i + \a_{i+1} + \cdots + \a_j \ ,
~~~\a_{i,j} ~=~ \a_i  + \a_j\ , ~~~i < j \ ,&\nota {} \cr
&\a_{ij,k} ~=~ \a_{k,ij} ~=~\a_i + \a_{i+1} +\cdots + \a_j +\a_k\ , ~~~i< j \ , \cr
&\a_{ij,km} ~=~ \a_i + \a_{i+1} +\cdots + \a_j +\a_k+ \a_{k+1} +\cdots +\a_m\ , ~~~i< j \ , ~~k<m \ , \cr
&\a_{ij,km,4} ~=~ \a_i + \a_{i+1} + \cdots + \a_j +\a_k+ \a_{k+1} +
\cdots +\a_m+\a_4\ , ~~~i< j \ , ~~k<m \ , \cr}$$
i.e., the non-compact roots will be written as:
\eqna\satsevyy
$$\eqalignno{&\a_7 \ , ~~\a_{67}\ , ~~\a_{57}\ , ~~\a_{47}\ , ~~\a_{37}\ , ~~\a_{1,37}\ ,&\satsevyy {a}\cr
&\a_{2,47}\ , ~~  \a_{27}  \ , ~~
\a_{17}  \ , ~~  \a_{27,4}  \ ,  ~~  \a_{17,4}  \ ,
~~  \a_{27,45}  \ , ~~ &\satsevyy {b}\cr& \a_{17,34}  \ , ~~  \a_{17,45}  \ , ~~  \a_{27,46}
  \ ,~~  \a_{17,35}  \ , ~~  \a_{17,46}  \ ,
~~  \a_{17,36}  \ ,\cr&
 \a_{17,35,4}  \ , ~~  \a_{17,25,4}  \ ,
~~  \a_{17,36,4}  \ , ~~  \a_{17,26,4}  \ ,\cr&
 \a_{17,36,45}  \ ,
~~  \a_{17,26,45}  \ ,
~~  \a_{17,26,45,4}  \ ,
~~  \a_{17,26,35,4}  \ ,
~~  \a_{17,16,35,4}    ~=~ \tilde{\a}\ ,
}$$
where the first six roots in \satsevyy{a} are from the $sl(7)$ subalgebra, and the 21 in \satsevyy{b}
are those from  \satsevy{}.

Further, we give the correspondence between the signatures $\chi$
and the highest weight $\L$. The connection is through the Dynkin
labels:    \eqn\dynk{ m_i ~\equiv~ (\L+\r,\a^\vee_i) ~=~ (\L+\r, \a_i )\ , \quad
i=1,\ldots,7,} where ~$\L = \L(\chi)$, ~$\r$ is half the sum of the
positive roots of ~$\cg^\bac$, ~$\a_i$~ denotes the simple roots of
~$\cg^\bac$.  The explicit connection is: \eqn\rela{ n_i =
m_i \ , \quad  c ~=~ -\ha (n_\ta + n_7) ~=~  -\,\ha(   2n_1+2n_2 +
3n_3 + 4n_4 + 3n_5 + 2n_6  + 2n_7)   }

We shall use also   the so-called Harish-Chandra parameters:
\eqn\dynhc{ m_\b \equiv (\L+\r, \b )\ ,} where $\b$ is any positive
root of $\cg^\bac$. These parameters are redundant, since obviously they
are expressed in terms of the Dynkin labels, however,   some
statements are best formulated in their terms.\foot{Clearly, both the Dynkin and
Harish-Chandra labels have their origin in the BGG reducibility condition \bggr.}

There are several types of multiplets: the main type, which contains
maximal number of ERs/GVMs, the finite-dimensional and the discrete
series representations, and some reduced types of multiplets.

In the next Section we give the main type of multiplets and the main reduced type.

\newsec{Multiplets}

\subsec{Main type of multiplets}

\nt
The multiplets of the main type are in 1-to-1 correspondence with
the finite-dimensional irreps of ~$E_7\,$, i.e., they will be
labelled by  the seven positive Dynkin labels    ~$m_i\in\bbn$. As we mentioned,
it turns out that each such multiplet contains 56 ERs/GVMs whose
signatures can be given in the following pair-wise manner:
\eqna\tabl
$$\eqalignno{
&\chi_0^\pm ~=~ \{\, (
 m_1,
 m_2,
 m_3,
 m_4,
 m_5,
 m_6)^\pm\,;\,\pm\ha(m_\ta   + m_7)\,\} &\tabl{}\cr&
\chi_{a}^\pm ~=~ \{\, (
 m_1,
 m_2,
 m_3,
 m_4,
 m_5,
 m_{67})^\pm\,;\,\pm\ha(m_\ta   - m_7 )\,\} \cr&
\chi_{b}^\pm ~=~
\{\, (
 m_1,
 m_2,
 m_3,
 m_4,
 m_{56},
 m_7)^\pm\,;\,\pm\ha(m_\ta   - m_{67} )\,\} \cr&
\chi_{c}^\pm ~=~ \{\, (
 m_1,
 m_2,
 m_3,
 m_{45},
 m_6,
 m_7 )^\pm\,;\,\pm\ha(m_\ta   - m_{57} )\,\} \cr&
\chi_{d}^\pm ~=~ \{\, (
 m_1,
 m_{2,4},
 m_{34},
 m_5,
 m_6,
 m_7 )^\pm\,;\,\pm\ha(m_\ta   - m_{47} )\,\} \cr&
\chi_{e}^\pm ~=~ \{\, (
 m_1,
 m_4,
 m_{24},
 m_5,
 m_6,
 m_7 )^\pm\,;\,\pm\ha(m_\ta   - m_{2,47} )\,\} \cr&
\chi_{e'}^\pm ~=~ \{\, (
 m_{1,3},
 m_{24},
 m_4,
 m_5,
 m_6,
 m_7)^\pm\,;\,\pm\ha(m_\ta   - m_{37} )\,\} \cr&
\chi_{f}^\pm ~=~ \{\, (
 m_{1,3},
 m_{34},
 m_{2,4},
 m_5,
 m_6,
 m_7)^\pm\,;\,\pm\ha(m_\ta   - m_{27} )\,\} \cr&
\chi_{f'}^\pm ~=~ \{\, (
 m_3,
 m_{14},
 m_4,
 m_5,
 m_6,
 m_7)^\pm\,;\,\pm\ha( m_\ta - m_{1,37} )\,\} \cr&
\chi_{g}^\pm ~=~ \{\, (
 m_{1,34},
 m_3,
 m_2,
 m_{45},
 m_6,
 m_7)^\pm\,;\,\pm\ha(m_\ta - m_{27,4} )\,\} \cr&
\chi_{g'}^\pm ~=~
\{\, (
 m_3,
 m_{1,34},
 m_{2,4},
 m_5,
 m_6,
 m_7)^\pm\,;\,\pm\ha(m_\ta - m_{17})\,\} \cr&
\chi_{h}^\pm ~=~ \{\, (
 m_{1,35},
 m_3,
 m_2,
 m_4,
 m_{56},
 m_7)^\pm\,;\,\pm\ha( m_\ta - m_{27,45} )\,\} \cr&
\chi_{h'}^\pm ~=~ \{\, (
 m_{34},
 m_{1,3},
 m_2,
 m_{45},
 m_6,
 m_7)^\pm\,;\,\pm\ha(m_\ta - m_{17,4})\,\} \cr&
\chi_{j}^\pm ~=~ \{\, (
 m_{1,36},
 m_3,
 m_2,
 m_4,
 m_5,
 m_{67})^\pm\,;\,\pm\ha( m_\ta - m_{27,46} )\,\} \cr&
\chi_{j'}^\pm ~=~ \{\, (
 m_{35},
 m_{1,3},
 m_2,
 m_4,
 m_{56},
 m_7)^\pm\,;\,\pm\ha( m_\ta - m_{17,45}  )\,\} \cr&
\chi_{j''}^\pm ~=~ \{\, (
 m_4,
 m_1,
 m_2,
 m_{35},
 m_6,
 m_7)^\pm\,;\,\pm\ha( m_\ta - m_{17,34} )\,\} \cr&
\chi_{k}^\pm ~=~ \{\, (
 m_{1,37},
 m_3,
 m_2,
 m_4,
 m_5,
 m_6)^\pm\,;\,\pm\ha( m_\ta - m_{27,46}  )\,\} \cr&
\chi_{k'}^\pm ~=~ \{\, (
 m_{36},
 m_{1,3},
 m_2,
 m_4,
 m_5,
 m_{67})^\pm\,;\,\pm\ha( m_\ta - m_{17,46}   )\,\} \cr&
\chi_{k''}^\pm ~=~ \{\, (
 m_{45},
 m_1,
 m_2,
 m_{34},
 m_{56},
 m_7)^\pm\,;\,\pm\ha( m_\ta - m_{17,35} )\,\} \cr&
\chi_{\ell}^\pm ~=~ \{\, (
 m_{37},
 m_{1,3},
 m_2,
 m_4,
 m_5,
 m_6)^\pm\,;\,\pm\ha\, m_{25,34}  \,\} \cr&
\chi_{\ell'}^\pm ~=~ \{\, (
 m_{46},
 m_1,
 m_2,
 m_{34},
 m_5,
 m_{67})^\pm\,;\,\pm\ha\, m_{2,45,4}\,\} \cr&
\chi_{\ell''}^\pm ~=~ \{\, (
 m_5,
 m_1,
 m_{2,4},
 m_3,
 m_{46},
 m_7)^\pm\,;\,\pm\ha\, m_{2,56} \,\} \cr&
\chi_{m}^\pm ~=~ \{\, (
 m_{47},
 m_1,
 m_2,
 m_{34},
 m_5,
 m_6)^\pm\,;\,\pm\ha\, m_{2,45,4} \,\} \cr&
\chi_{m'}^\pm ~=~ \{\, (
 m_{56},
 m_1,
 m_{2,4},
 m_3,
 m_{45},
 m_{67})^\pm\,;\,\pm\ha\, m_{2,5}  \,\} \cr&
\chi_{m''}^\pm ~=~ \{\, (
 m_5,
 m_1,
 m_4,
 m_3,
 m_{2,46},
 m_7)^\pm\,;\,\pm\ha( m_{56} - m_2  )\,\} \cr&
\chi_{n}^\pm ~=~ \{\, (
 m_{57},
 m_1,
 m_{2,4},
 m_3,
 m_{45},
 m_6)^\pm\,;\,\pm\ha\, m_{2,5} \,\} \cr&
\chi_{n'}^\pm ~=~ \{\, (
 m_6,
 m_1,
 m_{2,45},
 m_3,
 m_4,
 m_{57})^\pm\,;\,\pm\ha(m_2 - m_5)\,\} \cr&
\chi_{n''}^\pm ~=~ \{\, (
 m_{56},
 m_1,
 m_4,
 m_3,
 m_{2,45},
 m_{67})^\pm\,;\,\pm\ha(m_5- m_2 )\,\} \cr
 }$$
where we have used for the numbers ~$m_\b ~=~ (\L(\chi)+\r,\b)$~
the same compact notation as in \nota{} for the roots $\b$, and the notation
~$(...)^\pm$~ employs the natural conjugation of the subalgebra ~$E_6\,$,
more precisely:
\eqn\conu{
\eqalign{ & (n_1,n_{2},n_{3},n_{4},n_{5},n_{6})^- ~=~ (n_1,n_{2},n_{3},n_{4},n_{5},n_{6})\cr
&(n_1,n_{2},n_{3},n_{4},n_{5},n_{6})^+ ~=~ (n_1,n_{2},n_{3},n_{4},n_{5},n_{6})^{E_6} ~\doteq~
(n_6,n_{2},n_{5},n_{4},n_{3},n_{1}) }}
Note that in \tabl{} the last entries with sign plus (resp. minus) are positive (resp. negative),
except in the cases ~$\chi_{m}^\pm\,$, $\chi_{n}^\pm\,$, $\chi_{n'}^\pm\,$.

The ERs in the multiplet are related by intertwining integral and
differential operators. The  integral operators were introduced by
Knapp and Stein \KnSt{}. In fact, these operators are defined for any ER,
not only for the reducible ones, the general action being:
\eqn\knast{\eqalign{ & G_{KS} ~:~ \cc_\chi ~ \llr ~ \cc_{\chi'} \ ,\cr
&\chi ~=~ \{\, n_1,n_{2},n_{3},n_{4},n_{5},n_{6} \,;\, c\, \} \ , \cr
&\chi' ~=~ \{\, (n_1,n_{2},n_{3},n_{4},n_{5},n_{6})^{E_6}\,;\, -c\, \} ~=~
\{\,n_6,n_{2},n_{5},n_{4},n_{3},n_{1} \,;\, -c\, \} }}
Obviously, the pairs in \tabl{}  are related by Knapp-Stein
integral operators, i.e.,
\eqn\ackin{  G_{KS} ~:~ \cc_{\chi^\mp} \lra \cc_{\chi^\pm} }
The action on the signatures is also called restricted Weyl reflection, since it represents
the nontrivial element of the 2-element restricted Weyl group which arises canonically
with every maximal parabolic subalgebra.\foot{Generically, the Knapp-Stein operators can be normalized
so that indeed ~$G_{KS} \circ G_{KS} = {\rm Id}_{\cc_\chi}\,$. However, this usually fails exactly for
the reducible ERs that form the multiplets, cf., e.g., \DMPPT.}

Matters are arranged so that in every multiplet only the ER with
signature ~$\chi_0^-$~ contains a finite-dimensional nonunitary
subrepresentation in  a finite-dimensional subspace ~$\ce$. The
latter corresponds to the finite-dimensional   irrep of ~$E_7$~ with
signature ~$\{ m_1\,, \ldots,\, m_7 \}$. The subspace ~$\ce$~ is
annihilated by the operator ~$G^+\,$,\ and is the image of the
operator ~$G^-\,$. The subspace ~$\ce$~ is
annihilated also by the \ido{} acting from ~$\chi_0^-$~ to ~$\chi_b^-$~
(more about this operator below).
 When all ~$m_i=1$~ then ~$\dim\,\ce = 1$, and in that case
~$\ce$~ is also the trivial one-dimensional UIR of the whole algebra ~$E_{7(-25)}$.
Furthermore in that case the conformal weight is zero: ~$d=9+c=9-\ha(m_{\ta}+m_7)_{\vert_{m_i=1}}=0$.

Analogously, in every multiplet only the ER with signature
~$\chi_0^+$~ contains holomorphic discrete series representation.
This is guaranteed by the criterion \Har{} that for such an ER
all Harish-Chandra parameters for non-compact roots must be negative, i.e.,
in our situation, ~$ n_\a ~<~ 0$, for ~$\a$~ from \satsevyy{}. [That this holds for
our ~$\chi_0^+$~ can be easily checked using the signatures \tabl{}.]

In fact, the Harish-Chandra parameters are reflected in the division of the ERs into
~$\chi^-$~ and ~$\chi^+$~: ~ for the ~$\chi^-$~ modules   less than half of the
27 non-compact Harish-Chandra parameters are negative (none for ~$\chi_0^-\,$, 13 for
~$\chi_n^-\,$, ~$\chi_{n'}^-\,$, ~$\chi_{n''}^-$), while for the
~$\chi^+$~ modules   more than half of the non-compact
27 Harish-Chandra parameters are negative (27 for ~$\chi_0^+\,$, 14 for
~$\chi_n^+\,$, ~$\chi_{n'}^+\,$, ~$\chi_{n''}^+$). In fact, as in the parenthesized examples,
it is true that the sum of the number of negative Harish-Chandra parameters for any pair
~$\chi^\pm$~ is equal to 27.

Note that the ER ~$\chi_0^+\,$~ contains also the conjugate anti-holomorphic discrete
series. The direct sum of the holomorphic and the antiholomorphic
representations are realized in  an invariant subspace ~$\cd$~ of
the ER ~$\chi_0^+\,$. That subspace is annihilated by the operator
~$G^-\,$,\ and is the image of the operator ~$G^+\,$.\nl
Note that the corresponding lowest weight GVM is infinitesimally
equivalent only to the holomorphic discrete series, while the
conjugate highest weight GVM is infinitesimally equivalent to the
anti-holomorphic discrete series.\nl
The conformal weight of the ER  ~$\chi_0^+$~ has the restriction ~$d = 9+c = 9 + \ha(m_\ta + m_7) \geq 18$.

The \idos\ correspond to non-compact positive roots of the root
system of ~$E_7$, cf. \Dob, i.e., in the current context, the roots
 given in   \satsevyy{}.

The multiplets are given explicitly in Fig. 1, where we use the
notation: ~$\L^\pm = \L(\chi^\pm)$.  Each \ido\ is
represented by an arrow accompanied by a symbol ~$i_{j...k}$~ encoding
the root ~$\b_{j...k}$~ and the number $m_{\b_{j...k}}$ which is involved in the BGG
criterion. This notation is used to save space, but it can be used due to
 the fact that only \idos\ which are
non-composite are displayed, and that the data ~$\b,m_\b\,$, which
is involved in the embedding ~$V^\L \lra V^{\L-m_\b,\b}$~ turns out
to involve only the ~$m_i$~ corresponding to simple roots, i.e., for
each $\b,m_\b$ there exists ~$i = i(\b,m_\b,\L)\in \{ 1,\ldots,7\}$,
such that ~$m_\b=m_i\,$. Hence the data ~$\b_{j...k}\,$,~$m_{\b_{j...k}}$~
is represented by ~$i_{j...k}$~ on the arrows.

The pairs ~$\L^\pm$~ are symmetric w.r.t. to the bullet in the middle
of the figure, and the dashed line separates the ~$\L^-$~ modules from
the ~$\L^+$~ modules.

\medskip

\nt
{\bf Interpretation:} ~~~Since the relation to the usual conformal algebras in $n$-dimensional
Minkowski space-time is one of our main motivations to study ~$E_{7(-25)}\,$, we would like to mention briefly  some analogies,
using an exposition that is written in the same context, cf. \Dobpeds, though the results are
contained in much older work \DMPPT,\TMP,\Dobc,\PeSo, see also \Dob.\nl
If we take the most basic example when the inducing ~$E_6$-representation in the ERs ~$\chi^\pm_0$~
is the trivial one: ~$(m_1,m_2,m_3,m_4,m_5,m_6) = (1,1,1,1,1,1)$, then the conformal fields represented by
the ERs ~$\chi^\pm_0$~ are scalar, while those represented by
the ERs ~$\chi^\pm_a$~ are 27-dimensional vectors. There are invariant differential operators
depicted on Fig. 1:
\eqna\idoss
$$\eqalignno{ \cd_{m_7,\a_7} ~:&~ \cc_{\chi^-_0} ~\lra ~ \cc_{\chi^-_a} \   &\idoss{a}\cr
\cd_{m_7,\a_{17,16,35,4}} ~:&~ \cc_{\chi^+_a} ~\lra ~ \cc_{\chi^+_0} \  &\idoss{b}\cr}$$
Both  are equations of order ~$m_7\,$.
When the last free parameter ~$m_7 = 1$~ then
the ER  $\chi^-_a$~ is the analog of the vector potential ~$A_\nu\,$,
while the ER ~$\chi^+_a$~ is the analog of the current ~$J_\nu\,$.
Then the equations in \idoss{} are linear and can be written as:
\eqna\curr
$$\eqalignno{&  \pd_\nu \,\phi ~=~ A_\nu \ , \qquad \phi\in \cc_{\chi^-_0} \ ,
~~ A\in \cc_{\chi^-_a} &\curr{a}\cr
&  \sum_{\nu=1}^{27}\ \pd^\nu \,J_\nu ~=~ \Phi \ , \qquad \Phi\in \cc_{\chi^+_0} \ ,
~~ J\in \cc_{\chi^+_a} &\curr{b}\cr
 }$$
When the parameter ~$m_7 > 1$, then the analogs of \idoss{} are also treated in the older
references cited above (for instance \idoss{b}   would be an equation of
partial conservation).\nl
In all cases, we stress that these are ~{\it invariant}~ differential equations, on- and off-shell.\nl
Naturally, this is only a glimpse in the analogies with the usual conformal case,
much more will be said elsewhere, \Dobp.~~\dia

\medskip

In the next Subsection we shall consider the main type of reduced multiplets.

\subsec{Main type of reduced multiplet}

\nt
The multiplets of reduced type R7 contain 42 ERs/GVMs and may be obtained formally from the main type by
setting ~$m_7 ~=~ 0$. Their signatures are given explicitly by:
\eqna\tabla
$$\eqalignno{
&\chi_0^\pm ~=~ \{\, (
 m_1,
 m_2,
 m_3,
 m_4,
 m_5,
 m_6)^\pm\,;\,\pm\ha m_\ta   \,\} &\tabla{}\cr&
\chi_{b}^\pm ~=~
\{\, (
 m_1,
 m_2,
 m_3,
 m_4,
 m_{56},
 0)^\pm\,;\,\pm\ha(m_\ta   - m_{6} )\,\} \cr&
\chi_{c}^\pm ~=~ \{\, (
 m_1,
 m_2,
 m_3,
 m_{45},
 m_6,
 0 )^\pm\,;\,\pm\ha(m_\ta   - m_{56} )\,\} \cr&
\chi_{d}^\pm ~=~ \{\, (
 m_1,
 m_{2,4},
 m_{34},
 m_5,
 m_6,
 0 )^\pm\,;\,\pm\ha(m_\ta   - m_{46} )\,\} \cr&
\chi_{e}^\pm ~=~ \{\, (
 m_1,
 m_4,
 m_{24},
 m_5,
 m_6,
 0 )^\pm\,;\,\pm\ha(m_\ta   - m_{2,46} )\,\} \cr&
\chi_{e'}^\pm ~=~ \{\, (
 m_{1,3},
 m_{24},
 m_4,
 m_5,
 m_6,
 0)^\pm\,;\,\pm\ha(m_\ta   - m_{36} )\,\} \cr&
\chi_{f}^\pm ~=~ \{\, (
 m_{1,3},
 m_{34},
 m_{2,4},
 m_5,
 m_6,
 0)^\pm\,;\,\pm\ha(m_\ta   - m_{26} )\,\} \cr&
\chi_{f'}^\pm ~=~ \{\, (
 m_3,
 m_{14},
 m_4,
 m_5,
 m_6,
 0)^\pm\,;\,\pm\ha( m_\ta - m_{1,36} )\,\} \cr&
\chi_{g}^\pm ~=~ \{\, (
 m_{1,34},
 m_3,
 m_2,
 m_{45},
 m_6,
 0)^\pm\,;\,\pm\ha(m_\ta - m_{26,4} )\,\} \cr&
\chi_{g'}^\pm ~=~
\{\, (
 m_3,
 m_{1,34},
 m_{2,4},
 m_5,
 m_6,
 0)^\pm\,;\,\pm\ha(m_\ta - m_{16})\,\} \cr&
\chi_{h}^\pm ~=~ \{\, (
 m_{1,35},
 m_3,
 m_2,
 m_4,
 m_{56},
 0)^\pm\,;\,\pm\ha( m_\ta - m_{26,45} )\,\} \cr&
\chi_{h'}^\pm ~=~ \{\, (
 m_{34},
 m_{1,3},
 m_2,
 m_{45},
 m_6,
 0)^\pm\,;\,\pm\ha(m_\ta - m_{16,4})\,\} \cr&
\chi_{j}^\pm ~=~ \{\, (
 m_{1,36},
 m_3,
 m_2,
 m_4,
 m_5,
 m_{6})^\pm\,;\,\pm\ha( m_\ta - m_{26,46} )\,\} \cr&
\chi_{j'}^\pm ~=~ \{\, (
 m_{35},
 m_{1,3},
 m_2,
 m_4,
 m_{56},
 0)^\pm\,;\,\pm\ha( m_\ta - m_{16,45}  )\,\} \cr&
\chi_{j''}^\pm ~=~ \{\, (
 m_4,
 m_1,
 m_2,
 m_{35},
 m_6,
 0)^\pm\,;\,\pm\ha( m_\ta - m_{16,34} )\,\} \cr&
\chi_{k''}^\pm ~=~ \{\, (
 m_{45},
 m_1,
 m_2,
 m_{34},
 m_{56},
 0)^\pm\,;\,\pm\ha( m_\ta - m_{16,35} )\,\} \cr&
 \chi_{\ell}^\pm ~=~ \{\, (
 m_{36},
 m_{1,3},
 m_2,
 m_4,
 m_5,
 m_{6})^\pm\,;\,\pm\ha( m_\ta - m_{16,46}   )\,\} \cr&
\chi_{m}^\pm ~=~ \{\, (
 m_{46},
 m_1,
 m_2,
 m_{34},
 m_5,
 m_{6})^\pm\,;\,\pm\ha\, m_{2,45,4}\,\} \cr&
\chi_{\ell''}^\pm ~=~ \{\, (
 m_5,
 m_1,
 m_{2,4},
 m_3,
 m_{46},
 0)^\pm\,;\,\pm\ha\, m_{2,56} \,\} \cr&
\chi_{m''}^\pm ~=~ \{\, (
 m_5,
 m_1,
 m_4,
 m_3,
 m_{2,46},
 0)^\pm\,;\,\pm\ha( m_{56} - m_2  )\,\} \cr&
\chi_{n}^\pm ~=~ \{\, (
 m_{56},
 m_1,
 m_{2,4},
 m_3,
 m_{45},
 m_6)^\pm\,;\,\pm\ha\, m_{2,5} \,\} \cr&
\chi_{n''}^\pm ~=~ \{\, (
 m_4,
 m_3,
 m_{2,45},
 m_1,
 m_6,
 m_{56})^\pm\,;\,\pm\ha(m_5 - m_2)\,\} \cr
}$$

Here the ER ~$\chi_0^+$~ contains limits of the (anti)holomorphic discrete series representations.
This is guaranteed by the fact that for this ER
all Harish-Chandra parameters for non-compact roots are non-positive, i.e.,
~$ n_\a ~\leq~ 0$, for ~$\a$~ from \satsevyy{}.
The conformal weight has the restriction ~$d = 9+c = 9 + \ha\,m_\ta \geq 17$.

There are other limiting cases, where there are zero entries for the first six ~$n_i$~ values.
In these cases the induction procedure would not use finite-dimensional irreps of the ~$E_6$~
subgroup. The corresponding ERs would not have direct physical meaning,
however, the fact that they are together with the physically meaningful ERs has important
bearing on the structure of the latter.

Altogether, the analysis of the Harish-Chandra parameters reveals the following. For any ER there is exactly one
Harish-Chandra parameter (counting all, not only the non-compact) that is zero. The compact ones are seen in the list above.
The non-compact are as follows:
\eqn\zerohc{\eqalign{ &\chi_0^- ~:~ n_7=0 , \quad \chi_0^+ ~:~ n_\ta=0 , \cr
&\chi_j^\pm \,,  ~\chi_{\ell}^\pm \,,  ~\chi_m^\pm \,,  ~\chi_n^\pm \,, ~ \chi_{n''}^\pm \,, \quad ~:~ n_{27,46}=0 .}}
As in the main type,   ~ for the ~$\chi^-$~ modules   less than half of the
27 non-compact Harish-Chandra parameters are negative (none for ~$\chi_0^-\,$, 13 for ~$\chi_{n''}^-\,$), while for the
~$\chi^+$~ modules ~-~ except  ~$\chi_{n''}^+\,$ ~-~  more than half of the non-compact
27 Harish-Chandra parameters are negative (26 for ~$\chi_0^+\,$, 14 for ~$\chi_n^+\,$). In fact,
it is true that for any pair ~$\chi^\pm$~ the sum of the number of negative Harish-Chandra parameters is equal to 26.

These multiplets are depicted on Fig. 2.
The Weyl-conjugated pairs ~$\L^\pm$~ are symmetric w.r.t. to the bullet in the middle
of the figure, and the dashed line separates the ~$\L^-$~ modules from
the ~$\L^+$~ modules. The fact that the pair ~$\chi_{n''}^-\,$, ~$\chi_{n''}^+\,$, sits on the dashed line
signifies the fact that for these two ERs the number of negative non-compact Harish-Chandra parameters
equals the number of positive non-compact Harish-Chandra parameters, and that equals 13.
Note also that the ten ERs for which holds ~$n_{27,46}=0$, cf. \zerohc{}, are situated on two conjugated lines.

There are many other types of reduced multiplets, and their study may be done as in the case
of ~$E_{6(-14)}\,$~ in \Dobesix, but for ~$E_{7(-25)}\,$~ it will need much  more space, so we leave it for
a future publication.

\newsec{Outlook}

\nt In the present paper we continued the programme outlined in \Dobinv\ on the example
of the non-compact group  $E_{7(-25)}\,$. Similar explicit descriptions are planned for the other
non-compact groups, in particular those  with  highest/lowest weight representations. We
plan also to extend these considerations   to the
supersymmetric cases   and also to the quantum group setting.
Such considerations are expected to be very useful    for
applications to string theory and integrable models, cf., e.g.,
\Witten.

In our further plans it shall be very useful that (as in \Dobinv) we
 follow a procedure in representation
theory in which \idos\ appear canonically \Dob{} and which procedure
has been generalized to the supersymmetry setting \DobPet,\Dobsusy{}
and to quantum groups \Dobqg.  (For   more references, cf. \Dobinv.)

\bigskip

\nt {\bf Acknowledgements.}

\nt The author would like to thank for hospitality the Abdus Salam
International Centre for Theoretical Physics, where part of the work
was done. The author was  supported in part by
  the European RTN network {\it ``Forces-Universe''} (contract
No.{\it MRTN-CT-2004-005104}), by Bulgarian NSF grant  {\it DO
02-257}, and by the Alexander von Humboldt Foundation in the
framework of the Clausthal-Leipzig-Sofia Cooperation.

\parskip=0pt
\listrefs
\np

\fig{}{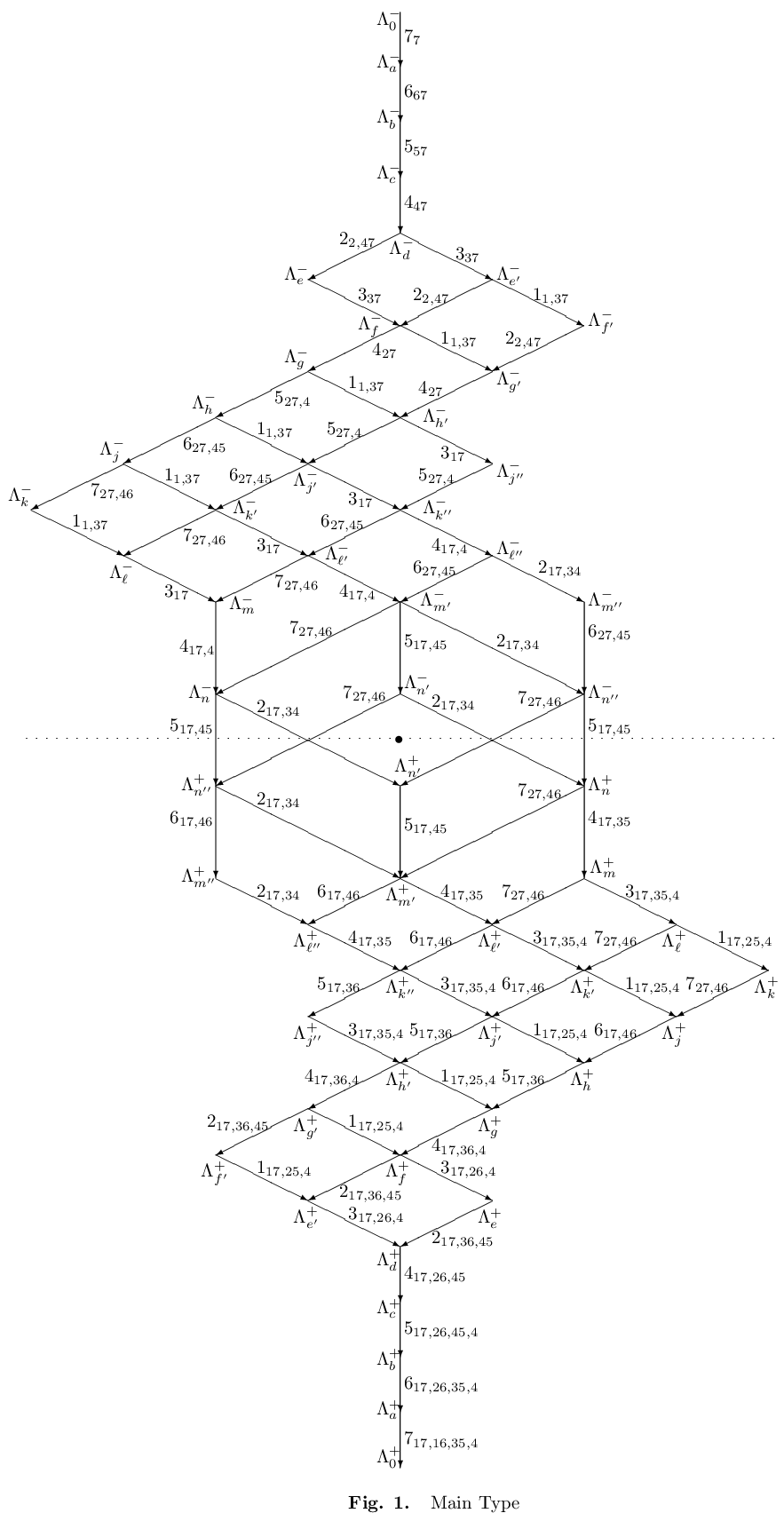}{10cm}

\np

\fig{}{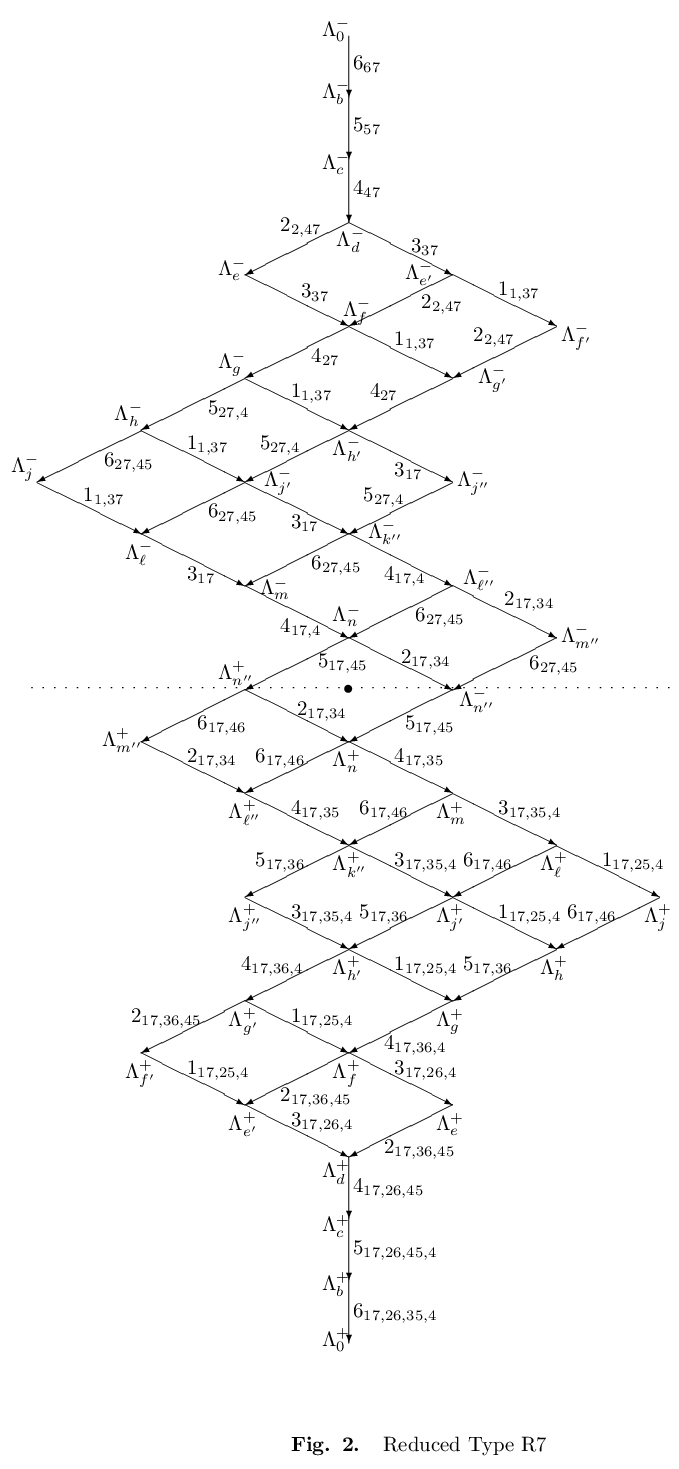}{7.7cm}

\np\end